\documentclass[a4paper,12pt]{article}
\usepackage[cp1251]{inputenc}
\usepackage[russian]{babel}
\usepackage{indentfirst}
\usepackage{amssymb,amsfonts}
\usepackage{graphicx}
\usepackage{amsmath}
\usepackage{amssymb,setspace}
\usepackage{afterpage}
\usepackage{longtable}
\usepackage{bm}
\usepackage[mathscr]{eucal}
\nonstopmode
\newcommand{\up}{\uparrow}
\newcommand{\dw}{\downarrow}

\newcommand{\lb}{\langle}
\newcommand{\rb}{\rangle}

\begin{document}
\begin{center}
\section*{Resonant excitation of the spin-wave current in hybrid nanostructures }

\end{center}

\begin{center}
 I. I. Lyapilin, M. S. Okorokov, V. V. Ustinov\\
 Institute of Metal Physics, UD RAS, Ekaterinburg,  Russia\\

email: Okorokovmike@gmail.com

\end{center}

{\small
Using the non-equilibrium statistical operator method (NSO), we have investigated the spin transport through the interface in a semiconductor/ferromagnetic insulator hybrid structure. We have analyzed the approximation of effective parameters, when each of the considered subsystems (conduction electrons, magnons, and phonons) is characterized by its effective temperature. We have constructed the macroscopic equations, describing the spin-wave current caused by both  resonantly excited spin system of conduction electrons and by an inhomogeneous thermal field in the ferromagnetic insulator.
}


\section*{Introduction}

Non-equilibrium statistical mechanics covers many problems of the interaction between a macroscopic system consisting of several sub-systems and external fields. In this case, the system’s non-equilibrium state depends on both  the external field energy absorbed by the subsystems per unit time and  the energy exchange rate between the subsystems, and on the energy leakage rate from the non-equilibrium subsystems toward a thermostat. The typical examples of these are the well-known Overhauser  \cite{Over} and Feher \cite{Feer} effects.  The latter are exhibited through observable deviations in interacting subsystems of a crystal as one of which (nuclear or electron) previously disturbed.
In spintronics, the spin pumping effect has been implemented in a paramagnetic/ferromagnetic insulator system \cite{Ts, Miz, Sai, Kaj}. The effect mentioned above is an analog of the Overhauser effect resulting to electron polarization. The saturation of ferromagnetic resonance in a localized spin subsystem causes the electrons to heat up. Otherwise speaking, their excess energy produces a deviation of the other subsystems from the equilibrium: the electron spins of the paramagnetic, phonons, etc. At the same time, the magnetic subsystem (localized spin subsystem of the ferromagnetic) transfers the spin angular momentum to the electronic subsystem of the paramagnetic material. It has been found that two types of  spin pumping are possible to exist. The first of these is resonance or \,"coherent". It involves the excitation of ferromagnetic (spin-wave) resonance in the localized spin subsystem. The paper \cite{An} demonstrates such a way of pumping in an experiment.  Another type is a non-coherent method of unbalancing either one or both the spin subsystems (subsystems of conduction electrons and localized spins) by external perturbations, for example, by thermal. Wherein, each of the spin subsystems can be described by its effective temperature. The different temperature relaxation ratios make it possible to realize both the spin pumping and the spin torque effect: the transfer of the spin angular momentum from the paramagnetic to the magnetic subsystem. The spin Seebeck effect (SSE) observed in $Ni_{81}Fe_{19}$ conducting crystals can serve as an example of the non-coherent spin pumping. Afterwards, the SSE could be observed in various materials, both semiconductors  and metallic ferromagnets  \cite{U3, U4, Jaw}. Besides, later the spin Nerst effect (or the thermal spin Hall effect), the spin Peltier effect, and others have been discovered \cite{ U5, Ad, Ma, Xu, Che, Du, Gra}.
The spin current is possible to be generated  by means of surface plasmons \cite{U1}.

Studying the SSE in a non-conducting magnet in the system of a non-magnetic conductor/magnetic insulator (N/F) $LaY_2 Fe_5O_{12}$  \cite{Uno} has shown that, against conducting crystals where the transfer of the spin angular momentum is due to band charge carriers, in non-conducting magnetic materials the spin Seebeck effect can be realized by exciting a localized spin system. For the SSE, the angular momentum transfer is driven by the spin-wave current (spin wave) underlain by excitations of the localized spin subsystem (magnons). Thus, unlike the conducting crystals, in the non-conducting magnet another spin current type - a spin-wave current can emerge. Since spin waves relax weakly enough, the spin-wave current propagates far greater distances than the electron spin current. This circumstance promises possible practical applications of the effect \cite{Taka, Bak}.

The work \cite{Kaj} investigates experimentally the spin-wave current. The detection as well as the generation of the spin current is a daunting task. The first detections of spin currents have been performed by indirect methods, by measuring the effects accompanied with the spin current generation. So, the work \cite{Kat} uses optical methods to measure the spin accumulation occurring on the lateral surfaces of the sample when generating the spin current in spin-Hall systems. Subsequently, the inverse spin Hall effect has been proposed as an electrical method of detecting spin currents \cite{Sai, Val, Kim}. The essence of this method is based on inducing a voltage by the spin current against the background of spin-orbit interaction \cite{Tak, Tak-1}. For the first time such a spin current detection method has been demonstrated in the work \cite{Sil} and has actually become the main method of detecting spin currents  \cite{An, Sek, Vil}.

The spin-wave current generation requires producing a non-equilibrium distribution of magnons, i.e. unbalancing the localized spin system. As has been noted above, this is possible to be done a variety of ways. Accompanied by the creation (or annihilation) of magnons, the interaction between the spin-polarized electrons and the localized spins at the interface (N/F) disturbs the magnetic subsystem. In the case of electron spin pumping, the effect is achieved by means of an alternating magnetic field under ferromagnetic resonance conditions. This way provides the spin current without transferring spin-polarized charge carriers through the interface in the hybrid structures. As such, the approach avoids the mismatch problem \cite{Ra, Zhu, Jon, Lou} that prevents to obtain high spin polarization values  by injection. The discovery of the spin Seebeck effect has shown that the spin current can be induced by thermal gradients. Thus, magnetic, electric, thermoelectric and quantum-relativistic (spin-orbit) interactions underlying various physical effects make the spin current generation possible.

 Both the spin Hall effect (SHE) and its inverse owe their origin mainly to the spin-orbit interaction (SOI). The latter couples the kinetic (translational) and spin subsystems of conduction electrons. Thus, the SOI is one of the possible channels to act on one of the subsystems via another, for example, on the spin subsystem of conduction electrons via the kinetic subsystem and vice versa. Due to the translational and spin motion locking, the quantum transitions cannot be conventionally divided into pure configurational (orbital) and pure spin ones. We can only talk about either predominantly configurational or predominantly spin transitions. But this circumstance significantly changes the conditions for the excitation of different transitions. Namely, the electrical component of an electromagnetic field initiates the spin transitions, and the magnetic component – the orbital ones. The spin-orbit interaction gives rise to the resonant electron transitions at frequencies being linear combinations of the cyclotron and Zeeman frequencies. Besides, such transitions can exist at the antinode of both electric and magnetic fields. Such resonance is known as the combined Rashba resonance \cite{Ras, Roi, Kal, Lya}. The powers absorbed by the electrons under saturation of the combined resonance (CR) and the paramagnetic resonance (PR) at the antinode of the electric field differ dramatically in magnitude. The former is much larger (by several orders of magnitude) than the latter. As to the combined resonance, a change in the average energy of the electron thermal motion (or their kinetic temperature) is chiefly explained by the fact that the kinetic degrees of freedom directly absorb the alternating electric field energy. However, as to the paramagnetic resonance, the kinetic degrees of freedom derive energy from the spin subsystem only indirectly, through the spin-orbital interaction.

 In this work we look at how the electric-dipole excitation of the electron spin subsystem of a semiconductor affects the spin-wave current generation in a non-conducting magnetic ferromagnetic material in the semiconductor/ferromagnetic insulator structure. The spin relaxation of conduction electrons is assumed to be due to the exchange interaction between the electrons and the localized spins located at the interface. Accompanied by the creation (or annihilation) of magnons, the inelastic spin-flip electron scattering when the electrons resonantly absorb energy from external fields (spin resonance saturation), causes the localized spin subsystem to non-monotonically deviate from its equilibrium state. Qualitatively, the effect can be explained as follows. The system is characterized by some average angular momentum having both electronic and magnetic components of an external magnetic field. The spin resonance saturation (the change in the spin temperature of the electron subsystem) changes the electronic component of the angular momentum. By virtue of the conservation law of angular momentum of the entire system, the magnetization of the magnetic subsystem must also be changed. This can be interpreted as a change in the spin temperature of the localized spin subsystem.

 The paper is organized as follows. The first part formulates the model at hand, contains the Hamiltonian of the system, and enters basic operators and their microscopic equations of motion. The second section involves constructing both the non-equilibrium entropy operator, accounting for the  perturbed system, and the Non-equilibrium Statistical Operator (NSO). The third section of the work covers analysis of macroscopic equations.

\section*{The  Hamiltonian}

 Consider a structure comprising two subsystems: a semiconductor and a ferromagnetic insulator (S/FI). Let them interact between each other and the lattice in crossed alternating electrical and magnetic fields. The external impact is assumed to give rise to combined (spin) resonance in the conduction electron subsystem.  Accompanied by the emission and absorption of magnons, the inelastic scattering of the conduction electrons by localized spins located near the interface, unbalances the localized spin system. As an additional mechanism of magnon scattering, the magnon-phonon interaction is  concerned. The system of conduction electrons in the semiconductor (S) should be viewed as a system consisting of kinetic and spin subsystems.
   The  Hamiltonian of the system can be represented as:
$$H = H_{SC}+H_{FI}+H_L +H_F.$$
Here
\begin{eqnarray}\label{1}
H_{SC} =  \int d{\bf x}\,H_e({\bf x}),\qquad\qquad\nonumber\\
  H_e({\bf x}) = \sum_j \{H_{e,j},\delta({\bf x}-{\bf x}_j)\} , \qquad   \,\,  H_{e,j} = H_{k,j} +  H_{s,j}+  H_{ks,j},\nonumber\\
H_k = \sum_j \frac{p_j^2}{2m},\,\, \qquad H_s = -\hbar\omega_s\,\sum_j\,s^z_j.\qquad
\end{eqnarray}
$ H_{ks}$ describes the interaction of the kinetic and spin degrees of freedom of the electrons. An expression for $ H_{ks}$  can be written in the general form
\begin{eqnarray}\label{2}
    H_{ks} = \sum_j f({\bf p}_j) S_j =  R\,\sum_j p_j^{\alpha_1}p_j^{\alpha_2}\ldots p_i^{\alpha_s}\,S_j^\beta,
  \end{eqnarray}
where $R$ is a constant depending on the spin-orbital interaction intensity.  $f({\bf p}_j)$   is a pseudo-vector whose components are a formula of order s of the kinetic impulse components $p_j^\alpha$. The integration is over the volume occupied by (S).   $s^z_j$  and $p^\gamma_j$  are the spin and momentum operator components of the j-th electron, respectively. $\omega_s=g_s\mu_0H/\hbar$ is the Zeeman precession frequency of free electrons in an external magnetic field directed along the z-axis ( $g_s,\,\mu_0$ are the effective electron spectroscopic splitting factor and the Bohr magneton, respectively); $\{ A, B \}=( A B + B A ) / 2$
 \begin{equation}\label{3}
H_{FI} =  \int d{\bf x}\, (H_m({\bf x}) +H_{sm}({\bf x}) )
\end{equation}
is the Hamiltonian of the localized spin subsystem.  $H_m({\bf x})$ is the energy density operator for the magnetic subsystem; it is of a sum of the exchange $H_{SS}(x)$ (over the nearest neighbors)  and Zeeman  energies $H_S(x)$.
\begin{equation}\label{2a}
H_{SS} = - J\, \sum_{j\delta}\,S_j S_{j+\delta},\,\,\, H_S = - \hbar\omega_m\sum_j\,S_j^z.
\end{equation}
J is the exchange integral, $\omega_m = g_m\mu_0 H/\hbar$.  $H_{ms}({\bf x})$   is the energy density operator of  interaction with the conduction electrons at the interface.
\begin{equation}\label{4}
 H_{sm} = -J_0\,\sum_j\int\,d{\bf x}\, {\bf s}({\bf x})\,{\bf S}({\bf R}_j)\,\delta({\bf x}-{\bf R}_j),
 \end{equation}
where $J_0$ is the exchange integral, $S({\bf R}_j)$ being the operator of the localized spin with the coordinate ${\bf R}_j$ at the interface. The integration in  (\ref{2}) is over the volume occupied by  $(FI)$.
 $H_L$   is the lattice Hamiltonian
 \begin{equation}\label{4a}
H_L = \int d{\bf x} \,(H_p({\bf x}) + H_{pm}({\bf x}) \,),
 \end{equation}
where $H_p({\bf x})$  is the energy density operator for the phonon subsystem. $H_{pm}({\bf x})$  is the energy density operator of interaction between the localized spins and phonons. $H_F $  is the interaction between the conduction electrons and the external alternating electrical field ${\bf E}(t)$.

 As a rule, the therms of the Hamiltonian $H_{ks}$  in a certain sense are small \cite{Ras}. In this case, to eliminate the interaction of the spin and kinetic degrees of freedom of the electrons in the linear approximation, we can perform an momentum-dependent canonical transformation of the Hamiltonian. This also modifies the rest of the Hamiltonian’s therms describing the interaction of the electrons with the lattice, and the electromagnetic field. As a result, we obtain a new Hamiltonian with autonomous subsystems (k)  and (s) , the electronic system effectively interacting with the electromagnetic field, which determines the resonant energy absorption. Such a transformation corresponds to gauge-invariant equations of motion for physical quantities under conditions, typical for combined resonance (CR). Suppose the canonical transformation of the Hamiltonian to be already made and the renormalized interaction with the alternating electric field has the form \cite{Lya}
\begin{eqnarray}\label{5}
\overline{H}_{ef}(t) = [{\bf r},\,T(p)]\,e\,{\bf E}(t) ,\qquad\qquad
T(p) = R\,\sum\frac{T^{\alpha_1,\ldots \alpha_s;\beta}}{\hbar\Omega_{\alpha_1,\ldots \alpha_s;\beta}}
\end{eqnarray}
where   $T(p)$  is the operator of the canonical transformation; $\Omega_{\alpha_1,\ldots \alpha_s;\beta}$  is a linear combination of the cyclotron  $\omega_0$ and the Zeeman  $\omega_s$ frequencies of electrons, this combination depends on the particular structure of the operator $T^{\alpha_1,\ldots \alpha_s;\beta}$  and the spin-orbital interaction constant.

To describe the non-equilibrium state of our system, we need calculate average energies of the subsystems (s) and (m) or their thermodynamically conjugate inverse effective temperatures $\beta_s, \beta_m$. Such a description corresponds to the case of establishing equilibrium inside each of the subsystems at a rate greater as compared to the energy exchange rate between them. We have earlier used it to analyze spin-thermal effects (the spin Seebeck effect) in hybrid structures \cite{L}. In general case, CR is possible to occur at frequencies being linear combinations of the cyclotron and the Zeeman frequencies. Then, both the kinetic and spin subsystems of conduction electrons are expected to be unbalanced. In the case of CR at the spin frequency, only the spin subsystem absorbs energy from the external alternating field, thereby the kinetic subsystem remains equilibrium. Thus, under combined (spin) resonance conditions, it is sufficient to discuss the evolution only two spin subsystem: the  conduction electrons spin subsystem and the localized spin subsystem. In the effective parameter approximation, the temperature $T_s$  characterizes the spin electron subsystem, the temperature $T_m$ – the localized spin subsystem of the ferromagnetic insulator. The equilibrium temperature $T$ corresponds to the phonon subsystem.

\section*{ The entropy operator}
To construct balance equations for average energies (or effective temperatures) of the subsystems, we employ the method of the non-equilibrium statistical operator (NSO) \cite{Zu, ZuK, L}. The entropy operator corresponding to a non-equilibrium state in terms of average density values can be written as
\begin{eqnarray}\label{6}
S(t)= \Phi(t) + \int d{\bf x}\,\{\beta\, (H_k({\bf x}) - \mu({\bf x,t})\, N({\bf x},t)  )+
 \beta_s({\bf x},t)\,(H_s({\bf x},t) + (1/2)H_{sm}({\bf x},t) ) \nonumber\\
+ \beta_m({\bf x},t) [H_m({\bf x},t) + (1/2)(H_{sm}({\bf x},t) +H_{pm}({\bf x},t))]+ \beta\,(H_p({\bf x},t) + (1/2) H_{pm}({\bf x},t) )\}\nonumber\\
  = S_0 + \delta S(t).\qquad
\end{eqnarray}
$\Phi(t)$  is the Massieu-Plank functional. $\beta_i(x,t), \quad i = s, m $  are local-equilibrium values of the inverse temperatures of the subsystems (s) and (m), respectively. $\mu({\bf x})$  is a local-equilibrium value of the chemical potential of electrons. $N({\bf x}) = \sum_i\,\delta({\bf x} -{\bf x}_i)$  is the electron number density operator. $S_0$  is the entropy of the equilibrium system with the Hamiltonian  $H$
$$S_0= \Phi_0 +\beta (H_e - \mu\, N) + \beta (H_m + H_p+ H_{sm}+ H_{mp}),$$
where  $\beta^{-1} = T$ is the equilibrium temperature of the system. The operator $\delta S(t) =  \int d{\bf x}\, \delta S({\bf x},t)$ describes the system deviation from its equilibrium state. Provided that the exchange interaction is the main mechanism of inelastic spin-electron scattering, we arrive at
\begin{eqnarray}\label{7}
\delta S(t)=\Delta\int d{\bf x}
\,\{ \delta\beta_s({\bf x},t) (H_s({\bf x},t) + (1/2)H_{sm}({\bf x},t) )- \beta\,\delta\mu({\bf x},t)\,N({\bf x}) \nonumber\\
+\delta\beta_m({\bf x},t) [H_m({\bf x},t) + (1/2)(H_{sm}({\bf x},t) + H_{mp}({\bf x},t))]\},
\end{eqnarray}
$$ \Delta A = A - \lb\,A\,\rb_0,\quad \lb\,\ldots\,\rb_0 = Sp\,(\ldots\rho_0).$$
$$\lb\,\ldots\,\rb^t = Sp\,(\ldots \rho(t)).$$
The non-equilibrium statistical operator $\rho(t)$  (NSO or the density matrix)  in the linear approximation in deviation from equilibrium can be written in the form \cite{ZuK}:
 \begin{eqnarray}\label{8}
\rho(t) =\rho_q(t) - \int\limits_{-\infty}^0\,dt_1\,e^{\epsilon t_1}\int\limits_0^1d\tau\, \rho_0^\tau\dot{S}(t+t_1,t_1)\rho_0^{1-\tau}.\qquad
\end{eqnarray}
Here $\rho_q(t)=\exp\{-S(t)\}$  is the quasi-equilibrium statistical operator,  $\dot{S}(t)$  is the entropy production operator
$$\dot{S}(t)=\delta\dot{S}(t) = \frac{\partial S(t)}{\partial t} +\frac{1}{i\hbar}[S(t), H].$$
A further algorithm for constructing the operator $\rho(t)$   reduces to finding the entropy production operator $\dot{S}(t)$. Commuting the operators $ H_s({\bf x}), N({\bf x}), H_m({\bf x})$ ,   with the Hamiltonian (\ref{1}), we come up with the operator equations of motion
 \begin{eqnarray}\label{9}
\dot{H}_s({\bf x}) &=& - {\bf \nabla}\,I_{H_s}({\bf x}) +(1/2) \dot{H}_{s(sm)}({\bf x})  + \dot{H}_{s(eF)}({\bf x}),\nonumber\\
\dot{N}({\bf x}) &=& - {\bf \nabla}\,I_{N}({\bf x}).
\end{eqnarray}
   Here
 \begin{eqnarray}\label{10}
 I_N({\bf x}) =  \frac{1}{m}\sum_j\{p_j,\delta( {\bf x}- {\bf x}_j)\},\qquad\qquad
I_{H_s}({\bf x}) = -\hbar\omega_s \frac{1}{m}\sum_js^z_j\{p_j,\delta( {\bf x}- {\bf x}^\alpha_j)\}
 \end{eqnarray}
are the particle flux and the Zeeman energy densities; $\dot{H}_{s(sm)}({\bf x})$  is the rate of change in local electron energy due to the interaction $H_{sm}({\bf x})$.   $\dot{H}_{s(eF)}({\bf x})$   determines the change in electron energy due to the interaction with the electrical field. Here  $\dot{A}_{\lambda(\lambda\gamma)}({\bf x})= (i\hbar)^{-1}[A_\lambda({\bf x})\,, H_{\lambda\gamma}]$.

Let us turn to the examination of the magnetic subsystem. Using the Holstein-Primakov method \cite{Uait}, the Hamiltonian of the localized spin subsystem can be represented through spin-wave (magnon) variables (using the creation $b^+_{\bf k}$ and annihilation $b_{\bf k}$ operators). Treating the magnon gas as free, we have
$$H_m = \sum_k \varepsilon(k) b^+_k\,b_k,\,\, \,\mbox{где}\,\,\, \varepsilon(k) = \frac{\hbar^2 k^2}{2 m^*}.$$
This expression can be interpreted as a sum of the energies of the quasiparticles-magnons having the quasi-momentum $ {\bf P}$ with their own effective mass $m^*$ and the magnetic moment \cite{Tur}. The equations of motion for the magnetic subsystem can be written in the form:
\begin{equation}\label{11}
\dot{H}_m({\bf x})= -\nabla\,I_{H_m}({\bf x})+(1/2)[\dot{H}_{m(sm)}({\bf x}) + \dot{H}_{m(pm)}({\bf x})].
 \end{equation}
Here
\begin{equation}\label{12}
 I_{H_m}({\bf x}) = - \hbar\omega_m\, I_{S^z}({\bf x})
\end{equation}
 is the magnon energy flux density. The rest of the terms in the right-hand side of the equation are responsible for the magnon scattering processes at the interface and by phonons. Then, the inelastic part of the exchange interaction can be written via the creation and annihilation operators for electrons and magnons:
\begin{equation}\label{13}
 H_{sm} = -J^*\,\sum_{k,k',q}\,\{b^+_q\,a^+_{k\up}\,a_{k'\dw} + b_q\,a^+_{k'\dw}\,a_{k\up} \}\, \delta_{{\bf k}',{\bf k}+{\bf q}},
 \end{equation}
where $a^+_{k\alpha}\,(a_{k\alpha})$   are the creation (annihilation) operators for electrons with a certain spin value $\alpha=\up,\,\dw$.

Ultimately, the equation of motion for the lattice subsystem has the form:
\begin{equation}\label{14}
\dot{H}_p({\bf x}) = -\nabla\,I_{H_p}({\bf x}) + (1/2)\dot{H}_{p(pm)}({\bf x}) .
 \end{equation}
Substituting the equations of motion found into the entropy production operator yields:
\begin{eqnarray}\label{15}
\dot{S}(t) = \Delta\int d{\bf x}
\,\{ - \beta\delta\mu({\bf x},t){\bf \nabla}\,I_{N}({\bf x}) +\beta \dot{H}_{s(sF)}({\bf x})
+ \delta\beta_s({\bf x},t)[ - {\bf \nabla}\,I_{H_s}({\bf x}) + \dot{H}_{s(sm)}({\bf x})  \nonumber\\
+\delta\beta_m({\bf x},t)[ -\nabla\,I_{H_m}({\bf x})+\dot{H}_{m(sm)}({\bf x}) + \dot{H}_{m(mp)}({\bf x})]\}.
\end{eqnarray}
Integrating by parts the terms containing the divergence of fluxes and discarding the surface integrals, we represent the entropy production operator as follows:
\begin{eqnarray}\label{16}
\dot{S}(t)=\Delta\int d{\bf x}\{I_N({\bf x})\beta\nabla\mu({\bf x},t) +I_{H_s}({\bf x})\nabla\beta_s({\bf x},t)\
 +I_{H_m}({\bf x})\nabla\beta_m({\bf x},t) +\beta \dot{H}_{s(F)} ({\bf x},t) \nonumber\\
 + \delta\beta_s({\bf x},t)\,(1/2)\dot{H}_{s(sm)} ({\bf x})\,+ \delta\beta_m({\bf x},t)(1/2)[\dot{H}_{m(sm)} ({\bf x}) +\dot{H}_{m(pm)}({\bf x})]\}.\quad
\end{eqnarray}
By expanding the quasi-equilibrium operator $\rho_q(t)=\exp\{-S(t)\}$   in powers of  $\delta S(t)$, we establish the linear relationship between the deviations of the thermodynamic coordinates and thermodynamic forces from their equilibrium values. Then, we have
 \begin{eqnarray}\label{17}
\delta\lb\,H_m({\bf x})\,\rb^t = -\!\!\int\!\! d{\bf x}'\delta\beta_m({\bf x}',t)(H_m({\bf x});H_m({\bf x}'))_0,\nonumber\\
\delta\lb\,H_s({\bf x})\,\rb^t = -\!\!\int\!\! d{\bf x}'\{\delta\beta_s({\bf x}',t)(H_s({\bf x});H_s({\bf x}'))_0
-\beta\delta\mu({\bf x}',t)(H_s({\bf x});N({\bf x}'))_0\},\nonumber\\
\delta\lb\,N({\bf x})\,\rb^t =\!\! -\int\!\! d{\bf x}'\{\delta\beta_s({\bf x}',t)(N({\bf x});H_s({\bf x}'))_0
-\beta\delta\mu({\bf x}',t)(N({\bf x});N({\bf x}'))_0\},
\end{eqnarray}
where
$$\delta \lb\, A \,\rb^t = \lb\, A \,\rb^t - \lb\, A \,\rb_0,$$
$$ (A,B)_0=\int\limits_0^1 d\lambda Sp\{A\rho^\lambda_0\Delta B\rho^{1-\lambda}_0\}.$$
Going over to the Fourier components of the spatial coordinates and taking into account that  $\lb\, N({\bf x}) \,\rb^t  = \lb\, N({\bf x}) \,\rb_0$, we get
\begin{eqnarray}\label{18}
\beta\delta\mu({\bf q},t) =  - \,\frac{(N({\bf q}),H_s(-{\bf q}))_0}{(N({\bf q}),N(-{\bf q}))_0}\,\delta\beta_s({\bf q},t),
\nonumber\\
\delta\lb\,H_s({\bf q})\,\rb^t = - \delta\beta_s({\bf q},t)(\hbar\omega_s)^2C_{zz}({\bf q}),
\end{eqnarray}
where
\begin{eqnarray*}
 C_{zz}(q) =(s^z({\bf q});s^z(-{\bf q}))_0  - \frac{(s^z({\bf q});N(-{\bf q}))_0(N({\bf q});s^z(-{\bf q}))_0}{(N({\bf q}); N(-{\bf q}))_0}.
\end{eqnarray*}
\begin{equation}\label{19}
\delta\lb\,H_m({\bf q})\,\rb^t = -\delta\beta_m({\bf q},t)(\hbar\omega_m)^2\,(S^z({\bf q});\,S^z(-{\bf q}))_0.
\end{equation}
Note that with help of  (\ref{17}), (\ref{18}), the entropy production operator  (\ref{16})  can appear as a functional of the Fourier components of the deviations of the thermodynamic coordinates from their equilibrium values.

\section*{ Macroscopic equations}

Averaging the operator equations (\ref{10}), (\ref{14}) by the NSO (\ref{8}), we can construct the macroscopic equations for the density of the spin magnetization of conduction electrons and localized spins: $ \delta m^z({\bf x},t) = g_s\mu_0\delta\lb\,s^z({\bf x})\,\rb^t$ and $\delta M^z({\bf x},t) = g_m\mu_0\delta\lb\,S^z({\bf x})\,\rb^t$ , respectively. We have
\begin{eqnarray}\label{23}
 \frac{\partial}{\partial t}\delta m^z({\bf x},t) =  - \nabla\lb\, I_{H_s}({\bf x},t)\,\rb^t+ \lb\,\dot{H}_{s(sm)}({\bf x},t)\,\rb^t
 + \lb\,\dot{H}_{s(ef)}({\bf x},t)\,\rb^t,\qquad\qquad\nonumber\\
  \frac{\partial}{\partial t}\delta M^z({\bf x},t) = - \nabla\lb\, I_{H_m}({\bf x},t)\,\rb^t + \lb\,\dot{H}_{m(sm)}({\bf x},t)\,\rb^t
 +\lb\,\dot{H}_{m(pm)}({\bf x},t)\,\rb^t\qquad\qquad
 \end{eqnarray}
Eqs. (\ref{23}) describe the change of the density of the spin magnetization of the electronic and magnetic subsystems due to the following processes: diffusion (the first summands in the right-hand sides of the equations), relaxation as a result of the exchange interaction between electrons with localized moments at the interface (the summands $\lb\,\dot{H}_{i(sm)}({\bf x},t)\,\rb^t ,i=s, m $), and energy absorption from an external electrical field by the spin subsystem of conduction electrons $\lb\,\dot{H}_{s(ef)}({\bf x},t)\,\rb^t\equiv Q_s({\bf x},t)$.

The summands $\lb\,\dot{H}_{m(pm)}({\bf x},t)\,\rb^t $  govern the magnon-phonon relaxation. In addition, the time and spatial dispersion effects should be also taken into account. These coefficients are given by the explicit expressions below
 \begin{eqnarray}\label{24}
\lb I_{H_s}^\alpha({\bf x},t)\rb^t\! =\!\! \int\!\!dx'\!\!\!\int\limits_{-\infty}^0\!\! dt_1e^{\epsilon t_1}\{ (I^\alpha_{H_s}({\bf x});I^\gamma_{H_s}({\bf x}',t_1))\nabla\beta_s^\gamma({\bf x}',\widetilde{t})\nonumber\\
+(I^\alpha_{H_s}({\bf x});I^\gamma_{N}({\bf x}',t_1))\beta\nabla\mu^\gamma({\bf x}',\widetilde{t})
+(I^\alpha_{H_s}({\bf x});I^\gamma_{H_m}({\bf x}',t_1))\,\nabla\beta_m^\gamma({\bf x}',\widetilde{t})\},
\end{eqnarray}
 \begin{eqnarray}\label{25}
\lb\, I_{H_m}^\alpha({\bf x},t)\,\rb^t\! =\!\! \int\!\!dx'\!\!\!\int\limits_{-\infty}^0\!\! dt_1e^{\epsilon t_1}\{ (I^\alpha_{H_m}({\bf x})\,;I^\gamma_{H_m}({\bf x}'t_1))\,\nabla\beta_m^\gamma({\bf x}',\widetilde{t}) \nonumber\\
+(I^\alpha_{H_m}({\bf x})\,;I^\gamma_{H_s}({\bf x}',t_1))\nabla\beta_s^\gamma({\bf x}',\widetilde{t})
+(I^\alpha_{H_m}({\bf x})\,;I^\lambda_{H_p}({\bf x}',t_1))\,\nabla\beta_m^\lambda({\bf x}',\widetilde{t})\},
\end{eqnarray}
  \begin{equation}\label{26}
\lb\dot{H}_{i(nm)}({\bf x},t)\rb^t\!\! =\!\! \int\!\!dx'\!\!\!\!\int\limits_{-\infty}^0\!\! dt_1e^{\epsilon t_1}\! (\dot{H}_{i(in)};\dot{H}_{i(in)}({\bf x}'\!,t_1))\delta\beta_j({\bf x}'\!,\widetilde{t}).
\end{equation}
\begin{equation}\label{27}
  Q({\bf x},t) =\!\! \beta\!\int\!\!dx'\!\!\!\int\limits_{-\infty}^0\!\! dt_1\,e^{\epsilon t_1} (\dot{H}_{s(eF)}({\bf x})\,;\dot{H}_{s(eF)}({\bf x' }\!,\!\widetilde{t})).
 \end{equation}
$\widetilde{t}=t+t_1;  i, j = s, m,\,\,n=m, p$. Eqs.  (\ref{23}) and the formulas for the kinetic coefficients (\ref{24}) - (\ref{27}) solve the problem of macroscopic description of the non-equilibrium spin subsystems in terms of average magnetization densities.

Next, we consider the case of stationary flow of energy between the subsystems. For this purpose, we average the equations  (\ref{23}) over time $t$ by the rule
 $$\overline{A(t)} = \epsilon\int_{-\infty}^0 dt \,A(t)\,e^{\epsilon t},\quad \epsilon\rightarrow+0,$$
 where
 $\overline{\partial_t A(t)} = 0, \overline{\delta\beta_i({\bf x},t+t_1)} = \delta\beta_i({\bf x}),\,\, i=s,m,p$.  Finally, the set of equations (\ref{23}) acquires the form
\begin{eqnarray}\label{28}
\delta\beta_{sm}({\bf x})\,L_{s(sm)}({\bf x}) + Q_s =0,\qquad\qquad\\
 \delta\beta_{ms}({\bf x})L_{m(sm)}({\bf x}) + \delta\beta_{mp}({\bf x})L_{m(mp)}({\bf x}) = 0.\nonumber\\
  \end{eqnarray}
 $\delta\beta_{ik} =\beta_i-\beta_k$  and
$$L_{i(jk)}({\bf x})\! =\!\!\int\!\! d{\bf x}'\!\!\! \int_{-\infty}^0 dt_1 e^{\epsilon t_1} (\dot{H}_{i(ik)}({\bf x});\dot{H}_{i(ik)}({\bf x}',t_1)).$$

Further, we proceed to the Fourier-representation in correlation functions and put that
$$ H_s({\bf x}) = \sum_q H_s({\bf q})\,e^{i{\bf q}{\bf x}},\quad H_s({\bf q}) = \sum_j H_{sj}\,e^{-i{\bf q}{\bf x}_j},$$

$$ I({\bf x}) = \sum_q I({\bf q})\,e^{i{\bf q}{\bf x}},\quad I({\bf q}) = \sum_j \{p_j/m, \,e^{-i{\bf q}{\bf x}_j}\},\ldots\mbox{и т. д.}$$
Now, we write down an expression for average power absorbed by the spin subsystem of conduction electrons under CR  $ Q_s =  \int d{\bf x}\overline{Q({\bf x},t)}$. According to \cite{Lya}, we have
\begin{equation}\label{29}
 Q_s\! = \! \beta \omega_s^2 R^2\sum_{{\bf q},\omega}\!\! |E^-(\omega)|^2\, \frac{\Gamma(q,\omega)\omega^2\,c_\pm({\bf q})}{(\omega-\omega_s)^2 +\Gamma^2(q,\omega)}.
  \end{equation}
Here  $c_\pm({\bf q}) =  (s^\pm({\bf q});s^\mp{\bf q}))_0,$
\begin{eqnarray}\label{30}
  \Gamma({\bf q},\omega) = \nu({\bf q},\omega) + q^\alpha\,q^\gamma\,D^{\pm}_{\alpha,\gamma}({\bf q},\omega),
  \end{eqnarray}
where $\nu({\bf q},\omega)$   is the known formula for the frequency of the transverse electron spin relaxation; it defines, for example, the line width of paramagnetic resonance \cite{Bik}
\begin{equation}\label{31}
\nu({\bf q},\omega)\! =\frac{1}{c_\pm({\bf q})}\,Re\!\!\int\limits_{-\infty}^0\!\! dt_1e^{(\epsilon-i\omega)t_1}(\dot{s}^+_{(sm)}({\bf q});\dot{s}^-_{(sm)}(-{\bf q},t_1)),
  \end{equation}
and $D^{\pm}_{\alpha,\gamma}({\bf q},\omega)$  is the diffusion tensor for the transverse spin magnetization components-
\begin{equation}\label{32}
D^{\pm}_{\alpha,\gamma}({\bf q},\omega) =\frac{1}{c_\pm({\bf q})}Re\!\!\int\limits_{-\infty}^0\!\! dt_1e^{(\epsilon-i\omega)t_1}(I^\alpha_{s^+}({\bf q})\,; I^\gamma_{s^-}(-{\bf q},t_1)),
  \end{equation}
The expressions for the kinetic coefficients  $\nu({\bf q},\omega),\,D^{\pm}_{\alpha,\gamma}({\bf q},\omega)  $  account for time and spatial dispersion and are suitable for both quantizing and classical magnetic fields. Previously, papers  \cite{Lya, Bik} have derived similar expressions for absorbed power in the homogeneous case.

Let us now look at the set of equations (\ref{23}) in the stationary case. We have
\begin{eqnarray}\label{33}
    (\beta_m\!-\!\beta_s)L_{m(sm)}({\bf q},\omega)\! +\! (\beta_m\!-\!\beta)L_{m(mp)}({\bf q},\omega)
     -q^\alpha\,q^\gamma\,D^{zz}_{\alpha,\gamma}({\bf q},\omega) = 0,\nonumber\\
  Q_s({\bf q},\omega) + (\beta_s\!-\!\beta) L_{s(sm)}({\bf q},\omega)  = 0   \qquad\qquad
 \end{eqnarray}
 where
\begin{equation}\label{34}
D^{zz}_{\alpha,\gamma}(q,\omega) = \frac{1}{C_{zz}({\bf q})}Re\!\!\int\limits_{-\infty}^0\!\!dt_1e^{(\epsilon-i\omega)t_1}(I^\alpha_{S^z}(q); I^\gamma_{S^z}(-q,t_1)),
  \end{equation}
and  $ C_{zz}({\bf q}) = (S^z({\bf q}),S^z(-{\bf q}))_0.$

  From (\ref{28}) it follows that  the expression for  spin-wave current in the ferromagnetic insulator is due to nonequilibrium magnon system and can be written  as
\begin{equation}\label{35}
\delta M^z({\bf q},\omega) =\,\frac{ \chi_0[q^\alpha\,q^\gamma\,D^{zz}_{\alpha,\gamma}({\bf q},\omega)+O_s({\bf q},\omega)(\omega_m/\omega_s)^2]}{L_{m(sm)}({\bf q},\omega)+L_{m(mp)}({\bf q},\omega)}\,\,
 \end{equation}
where  $\chi_0 = \beta(g_m\mu_0)^2(S^z(q);\,S^z(-q))$  is the static susceptibility of localized spins. As can be seen from the formula (\ref{35}), the spin-wave current depends on the frequency of an external field in a resonance manner. As for the correlation functions $L_{i(jk)}$  describing the relaxation processes, the papers \cite{Bik, Lya} have calculated them.

\section*{ Conclusions}

Using the non-equilibrium statistical operator method (NSO), we have investigated the spin transport through the interface in a semiconductor/ferromagnetic insulator hybrid structure. We have analyzed the approximation of effective parameters, when each of the considered subsystems (conduction electrons, magnons, and phonons) is characterized by its effective temperature. We have constructed the macroscopic equations describing the spin-wave current caused by both the resonantly exciting  spin subsystem of conduction electrons and  an inhomogeneous temperature field in the ferromagnetic insulator. Also, we have derived the generalized Bloch equations  describing the spin-wave current propagation in the insulator and taking into account the resonant-diffusion nature of the propagation of magnons and their relaxation processes.
We have shown that the spin-wave current excitation under combined resonance conditions bears a resonant nature.

The given work has been done as the part of the state task on the theme ”Spin” 01201463330 (project 15-17-2-17) with the support of the Ministry of Education of the Russian Federation (Grant 16-02-00044)


\begin{thebibliography}{99}

\bibitem{Over} A. Overhauser, Phys. Rev.  {\bf 92}, 411 (1953).
\bibitem{Feer} G. Feer Phys. Rev. Lett.  {\bf 3},135 (1959).
\bibitem{Ts}Y. Tserkovnyak, A. Brataas, G. E. W.  Bauer,  Phys. Rev. Lett. {\bf 88}, 117601 (2002).
\bibitem{Miz}S. Mizukami, Y. Ando, T. Miyazaki, Phys. Rev. B {\bf 66},104413 (2002).
\bibitem{Sai}E. Saitoh, M. Ueda, H. Miyajima, G. Tatara, G. Appl. Phys. Lett. {\bf 88},182509 (2006).
\bibitem{Kaj} Y. Kajiwara, K. Harii, S. Takahashi, J. Ohe, et.al., Nature {\bf 464}, 262 (2010).
\bibitem{An} K. Ando, S. Takahashi, J. Ieda, et.al., Nat. mat. {\bf 10}, 655  (2011).
\bibitem{U3}K. Uchida, K.et al., Nature {\bf 455},778 (2008).
\bibitem{U4}K. Uchida, K.et al., Nat. Mater. {\bf 9}, 894 (2010).
\bibitem{Jaw} C. M. Jaworski, J. Yang, S. Mack, et al., Nature Mater. 9, 898 (2010).
\bibitem{U5}K. Uchida, et al.,  Appl. Phys. Lett. {\bf 97}, 172505 (2010).
\bibitem{Ad} K. Uchida, J. Xiao, H. Adachi, et.al., Nature Materials {\bf 9}, 894 (2010).
\bibitem{Ma} Z.Ma, Sol. St. Comm. {\bf 150}, 505 (2010).
\bibitem{Xu} Xuele Liu, X.C. Xie, Sol. St. Comm. {\bf 150}, 471 (2010).
\bibitem{Che} S. Cheng, Y. Hing, Q. Sung, X. C. Xie, Phys. Rev. B. {\bf 78}, 045302 (2008).
\bibitem{Du} A. Durdal, J. Barnas, J. Phys. Cond. Mater. {\bf 24}, 275302 (2012).
\bibitem{Gra} L. Gravier, S. Serrano-Guisen, F. Reuse, et al., Phys. Rev. B. {\bf 73} 052410 (2006); {\bf 73}, 024419 (2006).
\bibitem{U1}K. Uchida,  H. Adachi, D. Kikuchi, S. Ito, et. al., Nat.Comm. {\bf 6}, 5910 (2015).
\bibitem{Uno} K. Uchida, T. Nonaka, T. Ota, et. al., Appl. Phys. Lett. {\bf 97}, 262504 (2010).
\bibitem{Taka} S. Takahashi, E. Saitoh, S. Maekawa, J. of Phys: Conf. Ser. {\bf 200}, 062030 (2010).
\bibitem{Bak} F. L. Bakker, A. Schlachter, J-P. Adam, et al., Phys. Rev. Lett. {\bf 105}, 136601 (2010).
\bibitem{Kat} Y. K. Kato, R. C. Myers, A. C. Gossard, et. al., Science {\bf 306}, 1910 (2004).
\bibitem{Val} Valenzuela, S. O. , Tinkham, M.  Nature {\bf 442}, 176 (2006).
\bibitem{Kim}  T. Kimura, Y. Otani, T. Sato, S. Takahashi, et.al.,  Phys. Rev. Lett. {\bf  98}, 156601 (2007).
\bibitem{Tak} S. Takahashi, S. Maekawa, Phys. Rev. Lett. {\bf 88}, 116601 (2002).
\bibitem{Tak-1} S. Takahashi, S. Maekawa, J. Phys. Soc. Jpn. {\bf 77}, 031009 (2008).
\bibitem{Sil} J. Sinova, D. Culcer, Q. Niu, N. A. Sinitsyn, T. Jungwirth, and A. H. MacDonald, Phys. Rev. Lett. {\bf 92}, 126603 (2004).
\bibitem{Sek} T. Seki, Y. Hasegawa, S. Mitani, et.al., Nature Mater.{\bf 7}, 125 (2008).
\bibitem{Vil} L. Vila, T. Kimura,  Y. C. Otani, Phys. Rev. Lett. {\bf 99}, 226604 (2007).
\bibitem{Ra} Rashba, E. I.  Phys. Rev. B  {\bf 62}, R17267 (2000).
\bibitem{Zhu}  Zhu, H. J. et al.  Phys. Rev. Lett. {\bf 87}, 016601 (1996).
\bibitem{Jon}  B. T. Jonker, B. Kioseoglou, G. Hanbicki, et.al.,  Nature Phys. {\bf 3}, 542 (2007).
\bibitem{Lou}  Lou, X. et al.  Phys. Rev. Lett. {\bf 96}, 176603 (2006).
\bibitem{Ras} E. I. Rashba, Sov. Phys.Usp. {\bf 84},557 (1964).
\bibitem{Roi} A. B. Roitsin,  Sov. Phys. Usp. {bf 14}, 766 (1972).
\bibitem{Kal}V. V. Kalashnikov, Theor. Mathem. Physics {\bf 18}, 76 (1974).
\bibitem{Lya}V. V. Kalashnikov, I.I.  Lyapilin, Theor. Mathem. Physics {\bf 18}, 194 (1974).
\bibitem{L}I. I. Lyapilin, M. S. Okorokov, V.V. Ustinov,  Phys.Rev.B  {\bf 91}, 195309 (2015).
\bibitem{Zu}D. N. Zubarev, Nonequilibrium statistical thermodynamics, M., Science, 1971.
\bibitem{ZuK} D. N. Zubarev, V. V. Kalashnikov, Theor. Mathem. Physics {\bf 1}, 137 (1969).
\bibitem{Uait} R. M. White,  Quantum theory of magnetism. (McGraw-Hill Book Company 1970).
\bibitem{Tur}{ S. V. Vonsovskii } Ferromagnetic Resonance. (Fiz.Mat.Lit., Moskau,1961).
\bibitem{Bik} H. M. Bikkin, V. P. Kalashnikov, Theor. Mathem. Physics {\bf 7}, 79 (1971).

 \end{thebibliography}
\end{document}